\documentclass[11pt, a4paper]{article}
\usepackage{moriond,epsfig}
\newcommand{\lmc}{l_{m_c}}

\begin{document}
\vspace*{4cm}
\title{PRECISE CHARM AND BOTTOM QUARK MASSES}

\author{JOHANN H. K\"UHN (1), MATTHIAS STEINHAUSER (1), CHRISTIAN STURM (2)} 
\address{(1) Institut f\"ur Theoretische Teilchenphysik,
   Universit\"at Karlsruhe, D-76128 Karlsruhe, Germany} 
\address{(2) Dipartimento di Fisica Teorica, Universit{\`a} di Torino, Italy, 
      INFN, Sezione di Torino, Italy}
\maketitle\abstracts{
New data for the total cross section
$\sigma(e^+e^-\to\mbox{hadrons})$ in the charm and bottom threshold
region are combined with an improved theoretical analysis, which
includes recent four-loop calculations, to determine the
short distance $\overline{\rm MS}$ charm and bottom quark masses.
The final result for the  $\overline{\rm MS}$-masses,
$m_c(3~\mbox{GeV})=0.986(13)$~GeV 
and
$m_b(10~\mbox{GeV})=3.609(25)$~GeV 
is consistent with but significantly more precise than a similar previous
study.}

%\noindent
%{\small¥{\it Keywords}: select here a few keywords for your article,
%from the A\&A thesaurus;\\
%visit {\tt ¥http://www.edpsciences.org/journal/index.cfm?edpsname=aa},
%and click on ``Instructions for authors''.}

The strong coupling constant and the quark masses are the fundamental input
parameters of the theory of strong interaction.
Quark masses are an essential input for the evaluation of weak decay
rates of heavy mesons and for quarkonium spectroscopy. 
Decay rates and branching ratios of a light Higgs boson, suggested by
electroweak precision measurements, depend critically on the masses of
the charm and bottom quarks, $m_c$ and $m_b$. Last not least, 
confronting the predictions for these masses with experiment is an 
important task for all variants of Grand Unified Theories.
To deduce the values in a consistent way from different experimental
investigations and with utmost precision is thus a must for current 
phenomenology.

A detailed analysis of $m_c$ and $m_b$ based on the ITEP sum rules 
\cite{Novikov:1977dq} has been
performed several years ago\cite{Kuhn:2001dm} and lead to 
$m_c(m_c)=1.304(27)$~GeV and $m_b(m_b)=4.191(51)$~GeV. 
During the past years new and more precise data for 
$\sigma (e^+e^-\to\mbox{hadrons})$ 
have become available in the low energy region,
in particular for the parameters of the charmonium and bottomonium
resonances. Furthermore, the error in the strong coupling constant
$\alpha_s(M_Z)=0.1189 \pm 0.0020$
has been reduced. Last not least, the 
vacuum polarization induced by massive quarks has recently been
computed in four-loop
approximation\cite{Chetyrkin:2006xg,Boughezal:2006px}; more 
precisely: its first derivative at $q^2=0$ has been evaluated, which
corresponds to the lowest moment of the familiar $R$-ratio.
With the help of the traditional integration-by-parts method in combination 
with Laporta's algorithm\cite{Laporta:1996mq,Laporta:2001dd} 
all  four-loop integrals were 
reduced to a small set of master integrals which were taken from  
Refs.~\cite{Schroder:2005va,Chetyrkin:2006dh}.
Based on these developments a new determination of the
quark masses has been performed in Ref.~\cite{Kuhn:2007vp}.

The extraction of $m_Q$ from low moments of the
cross section $\sigma(e^+e^-\to Q\bar{Q})$ exploits its sharp rise
close to the threshold for open charm and bottom production and the
importance of the contributions from the narrow quarkonium resonances. 
By evaluating the moments
\begin{eqnarray}
  {\cal M}_n &\equiv& \int \frac{{\rm d}s}{s^{n+1}} R_Q(s)
  \,,
  \label{eq::Mexp}
\end{eqnarray}
with low values of $n$, the long distance contributions are averaged
out and ${\cal M}_n$ involves short distance physics only, with a
characteristic scale of order $E_{\rm threshold}=2 m_Q$.
Through dispersion relations the moments are directly related to
derivatives of the vacuum polarization function at
$q^2=0$,
\begin{eqnarray}
  {\cal M}_n &=& \frac{12\pi^2}{n!}
  \left(\frac{{\rm d}}{{\rm d}q^2}\right)^n
  \Pi_Q(q^2)\Bigg|_{q^2=0}
  \,,
\label{eq::Mexp1}
\end{eqnarray}
which can be evaluated in perturbative QCD (pQCD). 

The narrow charmonium resonances $J/\Psi$, $\Psi(2S)$ and the
higher excitations will obviously contribute to the moments. Open
charm production exhibits a sharp rise, nearly like a step function.
Beyond the $\Psi(3770)$-resonance a few oscillations are observed which 
quickly level out into a fairly flat energy dependence.
Around and above approximately  5~GeV the cross section is well
approximated by pQCD and, furthermore, 
mass terms can be considered as small 
corrections\cite{Chetyrkin:1996ia,Chetyrkin:2000zk}. 
The sensitivity to $m_Q$ is, therefore, concentrated on the small region
from $J/\Psi$ up to approximately 5~GeV.

\begin{figure}[t]
  \begin{center}
    \begin{tabular}{c}
      \leavevmode
      \epsfxsize=.8\textwidth
      \epsffile{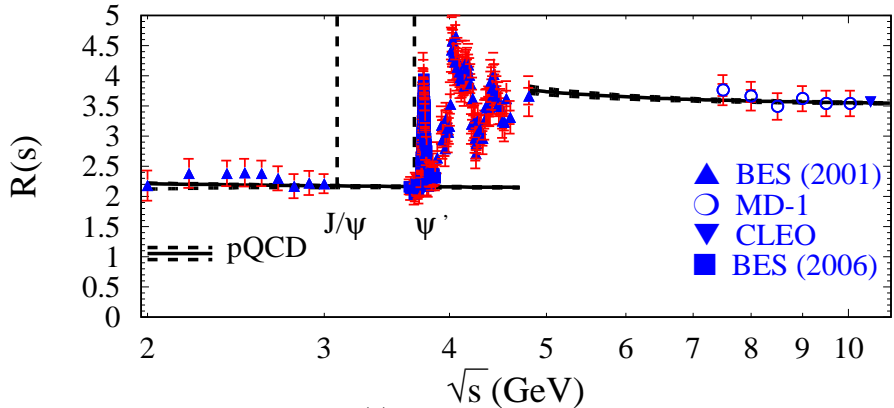}
      \\[-2em]
    \end{tabular}
  \end{center}
  \caption{\label{fig::R}$R(s)$ in 
    the charm threshold region. 
          }
\end{figure}

We therefore distinguish three energy regions: 
First, the region of the narrow resonances $J/\Psi$ and $\Psi(2S)$,
second, the ``charm threshold region''  starting from the $D$-meson threshold
at 3.73~GeV up to approximately 5~GeV, where the cross section exhibits rapid
variations and, third, the continuum region where pQCD and local duality are
expected to give reliable predictions. 
For the threshold region
we use the data from the BES collaboration \cite{Bai:2001ct,Ablikim:2006mb},
shown in Fig.~\ref{fig::R}
together with data from MD-1~\cite{Blinov:1993fw} and
CLEO~\cite{Ammar:1997sk}. 
Evidently pQCD 
provides an excellent description of all the data in the continuum region.
The description of the  perturbative continuum includes the complete 
mass dependence up to ${\cal O}(\alpha_s^2)$ plus the dominant mass dependent 
${\cal O}(\alpha_s^3)$ terms \cite{Kuhn:2007vp}   
which were used to  extrapolate $R_{uds}$ from the region
below charm threshold up to 4.8~GeV.

In its domain of analyticity $\Pi_c(q^2)$ can be cast into the form
\begin{eqnarray}
  \Pi_c(q^2) &=& Q_c^2 \frac{3}{16\pi^2} \sum_{n\ge0}
                       \bar{C}_n \left( \frac{q^2}{4m_c^2} \right)^n
  \,,
  \label{eq::pimom}
\end{eqnarray}
where $m_c=m_c(\mu)$ is the $\overline{\rm MS}$ 
charm quark mass at the scale $\mu$.
The perturbative series for the coefficients $\bar{C}_n$ 
in order $\alpha_s^2$ was evaluated in
Ref.~\cite{Chetyrkin:1995ii},%Chetyrkin:1997mb}, 
the four-loop contributions
to $\bar{C}_0$ and $\bar{C}_1$ in
Refs.~\cite{Chetyrkin:2006xg,Boughezal:2006px}.
The coefficients depend on $\alpha_s$ and on the charm quark mass
through logarithms of the form $\lmc\equiv\ln(m_c^2(\mu)/\mu^2)$.
Combining Eqs. (\ref{eq::Mexp}), (\ref{eq::Mexp1}) and  (\ref{eq::pimom}),
the charm quark mass can be obtained:
\begin{eqnarray}
  m_c(\mu) &=& \frac{1}{2} 
  \left(\frac{\bar{C}_n}{{\cal M}_n^{\rm exp}}\right)^{1/(2n)} 
  \,.
  \label{eq::mc1}
\end{eqnarray}
In the charm threshold region (which includes $\Psi(3770)$)
we have to identify the contribution from
the charm quark, i.e. we have to subtract the parts
arising from the light $u$, $d$ and $s$ quark.
In the continuum region above $\sqrt{s}=4.8$~GeV data are sparse and
imprecise. On the other hand, pQCD provides reliable predictions for
$R(s)$. Thus in this region we replace data by the theoretical prediction.

\begin{figure}[t]
  \begin{center}
    \begin{tabular}{c}
      \leavevmode
      \epsfxsize=0.95\textwidth
      \epsffile{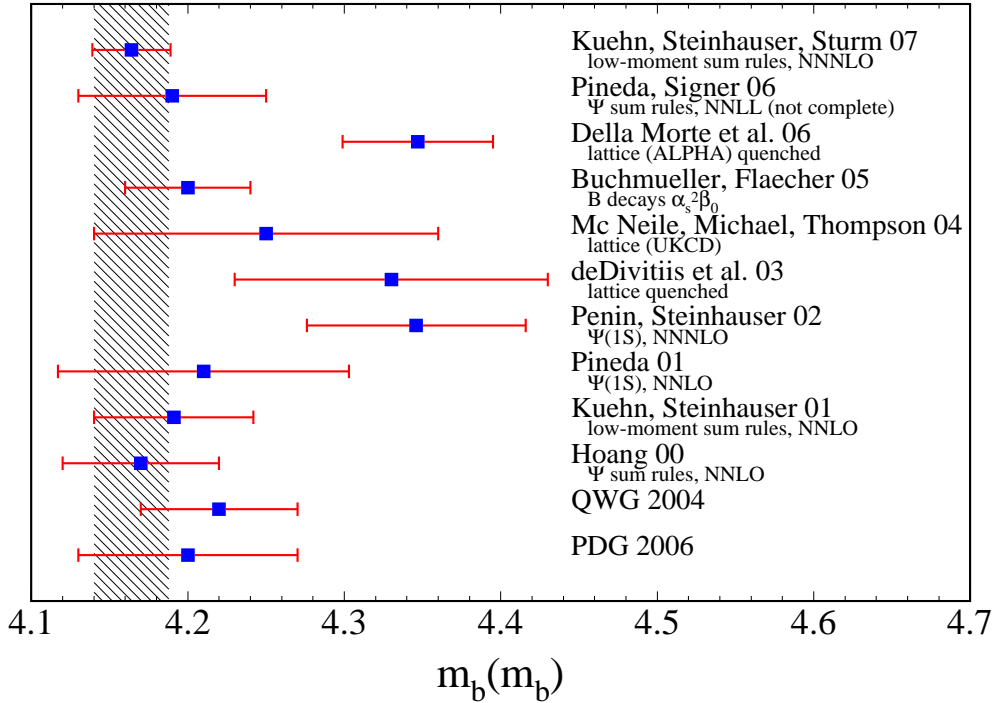}
    \end{tabular}
  \end{center}
  \vspace*{-2em}
  \caption{
    \label{fig::mb_compare}
    Comparison of recent determinations of $m_b(m_b)$ 
  }
\end{figure}

We use the results for the moments 
to obtain in a first step $m_c(3~\mbox{GeV})$.
The moment with $n=1$ is least sensitive to non-perturbative
contributions from condensates, to the Coulombic higher order effects, the
variation of $\mu$ and the parametric $\alpha_s$ dependence.
We therefore adopt 
\begin{eqnarray}
  m_c(3~\mbox{GeV}) &=& 0.986(13)~\mbox{GeV}
  \,,
  \label{eq::mc3final}
\end{eqnarray}
as our final result. 
Transforming this to the scale-invariant mass $m_c(m_c)$ one finds
$  m_c(m_c) = 1.286(13)~\mbox{GeV}$.
Using the three-loop 
relation~\cite{Chetyrkin:1999ys,Melnikov:2000qh}
between pole- and $\overline{\rm MS}$-mass this leads to
$M_c^{\rm(3-loop)} = 1.666~\mbox{GeV}$.

The same approach is also applicable to the determination of $m_b$. 
Just as in the charm case, a remarkable consistency and stability is
observed. For $n=1$ the error is dominated by the experimental input.
For $n=3$ we obtain $\pm 0.010$ from the experimental input,
$\pm 0.014$ from $\alpha_s$ and $\pm 0.006$ from the variation of $\mu$.
The three results based on $n=1,2$ and 3 are of comparable
precision. The relative size of the contributions from the threshold
and the continuum region decreases for the moments $n=2$ and 3. On the
other hand, the theory uncertainty estimated from the variation of
$\mu$ and the unknown four-loop contribution
is still acceptable. Therefore 
the result from $n=2$ is taken as the final answer,
\begin{equation}
  m_b(10~{\rm GeV} ) = 3.609(25)~\mbox{GeV}\,,
  \label{eq::mbten}
\end{equation}
corresponding to  $m_b(m_b) = 4.164(25)~\mbox{GeV}$ and
a pole mass of $M_b^{\rm(3-loop)} = 4.800~\mbox{GeV}$.
A comparison with a few selected determinations
is shown in Fig.~\ref{fig::mb_compare}.

For various applications, either related to $Z$-boson decays or in
connections to Grand Unified Theories (GUTs), the values of $m_b(\mu)$
at $M_Z$ and  $m_t(m_t) = 161.8 \pm 2.0~\mbox{GeV}$ (as derived from
$M_t=171.4\pm2.1~\mbox{GeV}$~\cite{Group:2006xn}) are of interest: 
\begin{equation}
  m_b(M_Z) = 2.834 \pm 0.019 \pm 0.017~\mbox{GeV}\,,~~~~~~~
  m_b(161.8) = 2.703 \pm 0.018 \pm 0.019~\mbox{GeV}
  \,.
\end{equation}
The first error reflects the combined error from
Eq.~(\ref{eq::mbten}) and the second one the uncertainty due to $\alpha_s$.
The ratio
$m_t(m_t)/m_b(m_t) = 59.8 \pm 1.3$
should be a useful input for Grand Unified Theories.

%- }}}
%- {{{ Acknowledgments:

\section*{Acknowledgments}

Work supported by
MIUR under contract
2001023713$\_$006, European Community's Marie-Curie Research
Training Network under contract MRTN-CT-2006-035505 `Tools and Precision
Calculations for Physics Discoveries at Colliders' and 
by the DFG through SFB/TR~9 `Computational Particle Physics'.

%- }}}

%- {{{ References:

\end{document}